\documentclass[reprint,amsmath,amssymb,aps]{revtex4-1}

\usepackage{graphicx}% Include figure files
\usepackage{dcolumn}% Align table columns on decimal point
\usepackage{bm}% bold math
\usepackage{appendix}
\usepackage[dvipsnames]{xcolor}
\begin{document}

\preprint{APS/123-QED}
%what is this?

\title{Effective Field Theory 
%treatment of 
for the Bound States and Scattering of\\
a Heavy Charged Particle and a Neutral Atom}

\author{Daniel Odell}
    \email{dodell@ohio.edu}
    \author{Daniel R. Phillips}%
    \email{phillid1@ohio.edu}
    \affiliation{
        Department of Physics \& Astronomy\\
        Institute of Nuclear \& Particle Physics\\
        Ohio University,
        Athens, OH 45701, USA
    }

\author{Ubirajara van Kolck}
    \affiliation{
        Universit\'e Paris-Saclay, CNRS/IN2P3, IJCLab,
        91405 Orsay, France
    }
    \affiliation{
        Department of Physics, University of Arizona,
        Tucson, AZ 85721, USA
    }

\date{\today}% It is always \today, today,

\begin{abstract}
We show the system of a heavy charged particle and a neutral atom can be described by a low-energy effective field theory where the attractive $1/r^4$  induced dipole potential determines the long-distance/low-energy wave functions. The $1/r^4$ interaction is renormalized by a contact interaction at leading order. Derivative corrections to that contact interaction give rise to higher-order terms. We show that this ``Induced-dipole EFT'' (ID-EFT) reproduces the $\pi^+$-hydrogen phase shifts of a more microscopic potential, the Temkin-Lamkin potential, over a wide range of energies. Already at leading order it also describes the highest-lying excited bound states of the pionic-hydrogen ion. Lower-lying bound states receive substantial corrections at next-to-leading order, with the size of the correction proportional to their distance from the scattering threshold. Our next-to-leading order calculation shows that  the three highest-lying bound states of the Temkin-Lamkin potential are well-described in ID-EFT. 
\end{abstract}

\maketitle

\section{Introduction \label{sec:Introduction}}

When a charged particle interacts with a neutral atom in an $S$ state at distances significantly larger than the Bohr radius, it experiences an attractive $1/r^4$ potential with a strength given by the atom's polarizability $\alpha$ \cite{VanVleck:1932}. 
%The strength of this potential is fixed by the polarizability of the hydrogen atom and the ratio of the charged particle's mass to the electron mass. 
The theory of singular potentials \cite{Frank:1971xx} therefore governs this situation, producing a particular pattern of bound states and scattering of the charged particle from the atom as a function of energy. 
Low-energy properties are given by $1/r^4$ dynamics, just as the $1/r^6$ potential determines the phase shifts and bound states of atom-atom systems for wave numbers of order one over the van der Waals length scale.
The analog of the van der Waals length scale is $\beta_4 \equiv \sqrt{\mu \alpha}$, where $\mu$ is the reduced mass of the atom-particle system.

Traditionally most attention has been devoted to electron scattering \cite{MM}, in particular the simplest case of the hydrogen ground state, whose polarizability is $\alpha_{\rm H}=9a_B^3/2$~\cite{VanVleck:1932,Mott:1948}, with $a_B$ the Bohr radius. Of particular current interest is the scattering of a heavier particle such as a negative muon ($\mu^-$), a positive pion ($\pi^+$), or a proton.
%--- scattering on the ground state of hydrogen
%This scenario is realized experimentally when the charged particle is, for %example, a negative muon ($\mu^-$), a positive pion ($\pi^+$), or a proton.
%for the $\mu^-$-hydrogen ion, pionic-hydrogen ions, and states of the H$_2^+$ ion where the electron is localized around one proton with the other proton far away. 
These systems have richer spectra than the electron case does, and rearrangement channels open up when the projectile is negatively charged, or positively charged and heavier than the proton.

Negatively charged heavy particles can also be captured in states with high orbital quantum number and cascade down to lower states, where they provide sensitive probes of nuclear properties \cite{Gotta:2004rq}. The quantum mechanics of their higher ionic levels is determined by the atomic polarizability and so it can also be explored with positively charged heavy particles.
%The $\pi^+$-hydrogen ion is of particular interest because of ongoing experimental efforts to measure its properties with precision. 
%{\bf add reference or remove?}
%All of these different systems have low-energy properties that are governed by $1/r^4$ dynamics, just as the $1/r^6$ potential determines the phase shifts and bound states of atom-atom systems for wave numbers of order one over the van der Waals length scale. The analog of the van der Waals length scale for these systems is $\beta_4 \equiv \sqrt{\mu \alpha_{H}}$ where $\mu$ is the reduced mass of the atom-particle system and $\alpha_H$ is the polarizability of hydrogen, which is 9/2 in atomic units~\cite{HydrogenPolarizability}. 

Such ions have $\beta_4 \gg a_B$.
%the Bohr radius. 
This scale separation enables an effective field theory (EFT) treatment of this problem. 
%EFT can compute the high-lying bound states and low-energy phase shifts of the $\pi^+$-hydrogen ion. This system is of particular interest because of ongoing experimental efforts to measure its properties with precision. 
%Effective Field Theory (EFT) 
EFT is a general tool that uses the separation of scales within a system, or class of systems, to make systematically improvable predictions  for observables within a well-defined energy window. 
(For 
an introduction to EFT, see Ref. \cite{Kaplan:1995uv}.) EFTs for 
singular potentials have been studied extensively~\cite{Beane:2001,Bawin:2003dm,Camblong:2003mb,Braaten:2004pg,Hammer:2005sa,PavonValderrama:2007nu,Long:2007vp,Bouaziz:2014wxa,Odell:2019wjq} for their relevance in atomic and nuclear physics~\cite{Hammer:2019poc}. They update and systematize the work of Case~\cite{Case:1953} and others on singular potentials in quantum mechanics \cite{Frank:1971xx}.
The $1/r^3$ potential is especially relevant as it is a leading piece of the interaction between two nucleons in Chiral EFT~\cite{Epelbaum:2008ga,Machleidt:2011zz,Hammer:2019poc,vanKolck:2020llt} in the limit that the pion mass is taken to zero~\cite{Bulgac:1997ji,Beane:2001bc}. 

The impact on observables of the $1/r^6$ potential in a ``van der Waals EFT" has been discussed recently~\cite{Odell:2021ryo}. Van der Waals EFT is similar to Quantum Defect Theory, which has been applied extensively to predict bound states and scattering in the situation that a $1/r^6$ potential determines the long-distance wave function of an atom-atom system~\cite{Gao1998SolutionsOT,GaoERE}.
Similar calculations have been done for the attractive $1/r^4$ potential~\cite{GaoInverseFourth}.

In contrast to the goals of EFT, where an accurate description of low-energy physics is primary, a semi-classical treatment is relevant in energy regimes where potential variations are ``slow'' in comparison to the wavelength of the wave function. The semi-classical predictions for attractive $1/r^n$ potentials were derived in Ref.~\cite{Flambaum:1999zza}.
At low energies, and certainly at threshold, the semi-classical approximation does not work well. Near threshold, the effective-range expansion (ERE) developed for finite-range potentials can be modified to account for the long-range tail; doing so generates additional terms that are non-analytic in energy. For $1/r^4$, these terms can be expanded ~\cite{OMalley:1961,PhysRev.125.1300} as a series in powers of the wave number in units of $\beta_4$, $k\beta_4$ times powers of $\beta_4/a_0$, where $a_0$ is the scattering length. The series starts with a linear correction to $k \cot \delta$. 

Instead of using either the ERE or a semi-classical approach, here we develop an EFT, Induced-dipole EFT (ID-EFT), that does not employ an expansion in $k \beta_4$, but instead expands  observables in powers of the ratio $a_B/\beta_4$ and the wave number in atomic units, $k a_B$ \footnote{For target atoms other than hydrogen, the size is, of course, larger than $a_B$}.
At leading order (LO) in this expansion the effect of the finite size of the atom on observables is captured in a smeared delta function. The Schr\"odinger equation is then solved for a potential consisting of that delta function and the $1/r^4$ induced-dipole potential. While the strength of the $1/r^4$ potential is fixed to be $\beta_4^2$, 
the strength of the delta-function piece depends on short-distance details of the atom-charged particle interaction and must be fit to one datum. Accuracy is improved in a next-to-leading-order (NLO) calculation, which introduces an additional piece of the short-distance potential, with an additional coupling constant, that must be fit to an additional datum. The same procedure is repeated at higher orders. In contrast to the ERE, which is limited to wave numbers $k$ for which $k \beta_4 \lesssim 1$, ID-EFT can handle wave numbers comparable to $1/\beta_4$. The ERE results if the induced-dipole potential is treated in perturbation theory as a higher-order effect.

As a specific example, we consider the bound states and phase shifts of the $\pi^+$-hydrogen ion. The spectrum is rich but scattering is not afflicted by open rearrangement channels. We use the pion-atom scattering length to fix the strength of the LO delta function. At NLO, the additional short-range parameter is fitted to the shallowest bound-state energy. Other energy levels in the system, as well as the scattering phase shifts, are then predicted by ID-EFT, up to corrections to each observable that have a fractional size $\sim k^4 a_B^4$, where $k$ is the characteristic wave number of the scattering or bound state. An important aspect of our calculation is that by carrying it out for several different choices of the delta-function smearing we can assess which of our observable predictions are independent of the details of this short-distance piece of the potential. 

Several studies, some of which treated  $\pi^+ H$  as a two-body system, and some of which treated it as a three-body system, have been conducted previously~\cite{Lazauskas:2002,Carbonell:2011,Lazauskas:2019ltg}.
%LazauskasPhD}. 
In the two-body treatment the pion-atom interaction was taken to be an analytical, parameter-free potential---the Temkin-Lamkin polarization potential~\cite{MM,Temkin:1959,Temkin:1961}---that superposes some short-range effects onto the $1/r^4$ tail. (The Temkin-Lamkin potential is also useful in complementing approximate solutions of the three-body system \cite{PhysRevA.42.6560}.) We take the Temkin-Lamkin potential's results for the $\pi^+ H$ system's bound-state energies and $S$-wave scattering phase shifts as data that allow us to assess the efficacy of ID-EFT for this system. We use those results as a laboratory to demonstrate the ability of ID-EFT to capture the low-energy portion of the rich spectrum and multi-faceted phase-shift behavior 
that results from the induced-dipole interaction. We are particularly interested in the fact that in many other applications of singular potentials --- for example in nuclear physics --- one contends with a single bound state, while the Temkin-Lamkin potential produces seven bound states in the $\pi^+ H$ system. It is therefore interesting to see how certain details of the EFT renormalization and calculation play out in this more complex situation.

Nevertheless, the Temkin-Lamkin treatment of $\pi^+ H$ is an approximation, and there are significant corrections to that approximation in a full three-body treatment ~\cite{Lazauskas:2002}. Having developed the basic ideas of ID-EFT in this work we intend to return to this problem in subsequent papers.   There we will instead use data from three-body treatments of the $\pi^+ H$ system as input to our EFT. 
 
The remainder of our paper is structured as follows. 
%The details of 
Our theoretical formulation 
%are 
is given in Sec.~\ref{sec:Theory}, with details of its implementation relegated to Apps. \ref{app:scattering_length}, \ref{app:branches} and \ref{app:NLOimplementation}.
The bound-state and scattering results at LO and NLO are presented and discussed in Sec.~\ref{sec:Results}, while details of extrapolations to small cutoffs are given in App. \ref{app:nlo_fit}.
Conclusions and future prospects are discussed in Sec.~\ref{sec:Conclusions}.

\section{Theory \label{sec:Theory}}

\subsection{Leading Order\label{sec:Theory_LO}}

% What do we keep fixed?
% g_LO(R)
% B_2^{(n)}(R)
% What's going on with the deeper states?
% kcotdelta

The LO interaction in ID-EDT takes the coordinate-space form
\begin{equation}
    \label{eq:lo_potential}
     V(r) = -\frac{C_4}{r^4} \;\rho(r;R) + g_{\rm LO}(R)\,\chi(r;R)~,
\end{equation}
which, for context, is input to the radial Schr\"odinger equation at energy $E$,
\begin{equation}
    \label{eq:schroedinger}
    \left[ -\frac{\hbar^2}{2\mu}\frac{d^2}{dr^2} + V(r) \right] u(r) = E u(r)~,
\end{equation}
where $\mu$ is the reduced mass of the charged particle-atom system.
The regulator functions $\rho(r;R)$ and $\chi(r;R)$ act, respectively, to overcome the $1/r^4$ potential and mimic the delta function at short distances.
Both interactions are regulated at the radius, $R$, 
related to the short-distance physics that 
we account for at LO through a contact interaction of strength $g_{\rm LO}(R)$.
The precise forms of $\rho(r;R)$ and $\chi(r;R)$ are not important, only that
\begin{align}
    \label{eq:rho_requirement}
    \lim_{r\rightarrow 0} \rho(r;R)/r^4 & =  0~, \\
    \lim_{R\rightarrow 0} \chi(r;R) & \propto \delta(r)~.
\end{align}
Here we take
\begin{align}
    \label{eq:v_lo_chi}
    \chi(r;R) & = e^{-(r/R)^4}~,\\
        \label{eq:v_lo_rho}
    \rho(r;R) & = \left[1-e^{-(r/R)^2}\right]^4~.
\end{align}
Once $g_{\rm LO}(R)$ is determined from one low-energy datum, LO is renormalized~\cite{Beane:2001}, namely, other low-energy observables converge as $1/R$ increases beyond the breakdown scale of the theory, $\Lambda_b\sim 1/a_B$.

The short-range interactions of ID-EFT capture the low-energy effects of physics at distances comparable to the 
%Bohr radius
atom's size. While 
%a three-body treatment shows that 
the asymptotic form of the potential is $1/r^4$, as the charged particle approaches its impact on the atom's distortion can no longer be accounted for solely by the polarizability \cite{Castillejo:1960}. Here, as an illustration of the method,
%At each value of $R$, 
we tune $g_{\rm LO}(R)$ 
at each value of $R$ to match the scattering length, $a_0$,
%$a_0 = -65$ a.u., 
obtained with the Temkin-Lamkin (TL) potential~\cite{Temkin:1959,Temkin:1961,MM},
\begin{equation}
    \label{eq:v_mm}
    V_{\rm TL}(r) = -\frac{e^2}{8\pi\epsilon_0} \frac{\alpha(r)}{4\pi\epsilon_0}\frac{1}{r^4}~,
\end{equation}
%\begin{equation}
%    \label{eq:v_mm}
%    V_{\rm TL}(r) = -\frac{e^2}{4\pi\epsilon_0} \frac{\alpha(r)}{4\pi\epsilon_0}\frac{1}{2r^4}~,
%\end{equation}
% {\bf How many minus signs?} Corrected.
where $e$ is the electron charge, $\epsilon_0$ is the vacuum permittivity, and
\begin{multline}
    \label{eq:alpha_r}
      \frac{\alpha(r)}{4\pi\epsilon_0} = 
      %\alpha_{\rm H}
      \frac{9}{2} a_B^3 
      \left\{1- e^{-2r/a_B}
      \left[ 1 + 2 r/a_B + 2(r/a_B)^2 
      \right. \right.\\ 
       \left. \left.
      + \frac{4}{3}(r/a_B)^3 
      + \frac{2}{3}(r/a_B)^4 
      + \frac{4}{27}(r/a_B)^5 
      \right]\right\}~.
\end{multline}
%\begin{multline}
%    \label{eq:alpha_r}
%      \frac{\alpha(r)}{4\pi\epsilon_0a_B^3} = \frac{9}{2} - \frac{2}{3} e^{-2r/a_B}\left[ (r/a_B)^5 +  \frac{9}{2}(r/a_B)^4 + \\ & 9(r/a_B)^3 + \frac{27}{2}(r/a_B)^2 + \frac{27}{2}(r/a_B) + \frac{27}{4} \right]~,
%\end{multline}
The TL potential is one of several semi-phenomenological polarization potentials \cite{PhysRevA.34.1810} that account for various effects associated with the interaction of the charged particle 
%(in this case the pion) 
with the full charge distribution of the 
%electron in the hydrogen 
atom. However, at distances $r \gg a_B$ only the $1/r^4$ piece of the potential survives. Matching to  Eq.~(\ref{eq:lo_potential}) we determine the length-scale associated with this piece of the potential as $\beta_4^2 \equiv 2\mu C_4/\hbar^2$, where
\begin{equation}
    \label{eq:c4}
    C_4 = \frac{e^2}{8\pi\epsilon_0} \lim_{r\rightarrow \infty} \frac{\alpha(r)}{4\pi\epsilon_0} 
    = \frac{1}{2} \left(\frac{e}{4\pi\epsilon_0}\right)^2 \alpha_{\rm H}
\end{equation}
%\begin{equation}
%   \label{eq:c4}
%    C_4 = \lim_{r\rightarrow \infty} -\frac{1}{2} \frac{e^2}{4\pi\epsilon_0} \frac{\alpha(r)}{4\pi\epsilon_0} = -\frac{1}{2} \left(\frac{e}{4\pi\epsilon_0}\right)^2 \alpha_{\rm H}
%\end{equation}
represents the strength of the $1/r^4$ potential and $\alpha_{\rm H}=9a_B^3/2$ is the polarizability of the hydrogen atom~\cite{VanVleck:1932,Mott:1948}.
%For the remainder of this article, we will express relevant quantities in atomic units (a.u.). In the case of the pion this produces $\beta_4=32.7$ a.u.
% {\bf Please check this number} Checked.
Details at short distances are not important in this application, and other potentials that curb the growth of $1/r^4$ would do as well. The %Temkin-Lamkin 
TL potential is a rich example as it supports many bound states when the charged particle is heavy.

$\beta_4$, and the scattering length, $a_0$, are the two physical scales that are inputs to ID-EFT at leading order. 
For the pion-hydrogen system, $\beta_4=32.7$ a.u. and the TL potential gives $a_0 = -65$ a.u.
Details of the determination of $a_0$ in the presence of a $1/r^4$ tail are given in App. \ref{app:scattering_length}.
In addition, 
since this is a local theory, there are, in principle, infinitely many values of $g_{\rm LO}$ that yield the desired scattering length $a_0$. Each such value produces a different number of bound states~\cite{Beane:2001}. Here we choose the branch of the implicit function $g_{\rm LO}(a_0)$ that corresponds to fourteen $S$-wave bound states, as described in App. \ref{app:branches}.
There are only seven states allowed by the 
%Temkin-Lamkin 
TL potential, but in order to study the renormalization of continuum and bound-state observables above the approximate breakdown scale of the theory, we choose a ``lower'' branch such that the repulsion of the LO counterterm is not numerically prohibitive.

With $\beta_4$ and $a_0$ fixed we can predict at LO all binding energies. We also compute the phase shifts
\begin{equation}
    \label{eq:phase_shifts}
    \delta^{({\rm LO})} = 
    \tan^{-1}\left(\frac{t^{({\rm LO})}}{1+it^{({\rm LO})}}\right)
\end{equation}
from 
\begin{equation}
    \label{eq:tLO}
    t^{({\rm LO})} = V_{\rm LO} + V_{\rm LO} G_0 t^{({\rm LO})}~,
\end{equation}
in a short-hand notation where an integral over the momentum
in the two-body propagator $G_0$ is implicit.

We note that the LO potential (\ref{eq:lo_potential}) contains no direct information on $\Lambda_b$.
%the scale $a_B$. 
As is typical in EFTs, quantitative information on the breakdown scale enters the calculation only in the context of higher-order corrections.

\subsection{Next-to-Leading Order \label{sec:Theory_NLO}}

% What do we keep fixed?
% g_NLO(R)
% B_2^{(n)}(R) – improvement!
%kcotdelta

At next-to-leading-order (NLO) the interaction is modified to
\begin{equation}
    \label{eq:nlo_potential}
    V(r) = -\frac{C_4}{r^4} \; \rho(r;R) 
    + \left(g_{\rm LO}(R) + E g_{\rm NLO}(R)\right) \chi(r;R)~,
\end{equation}
where we have chosen an energy-dependent NLO contribution to the short-distance potential.
%for reasons that will be discussed in Sec.~\ref{sec:Results}.
The new parameter $g_{\rm NLO}(R)$ is obtained from a second low-energy datum, and other observables converge as $R$ decreases as long as perturbation theory is employed. Corrections to the long-range polarization potential from the quadrupole polarizability and non-adiabatic contributions are $\propto 1/r^6$ \cite{PhysRevA.34.1810}. They should be included at this order if a phenomenological analysis were to be performed. Here, we do not consider these corrections as we are interested only in demonstrating the ability of ID-EFT to reproduce the low-energy effects of a given underlying potential (chosen to be the parameter-free TL). There is no difficulty of principle in including them along the lines of Ref. \cite{Long:2007vp}, where a $1/r^4$ correction to an attractive $1/r^2$ potential was considered. The additional long-range potential would give rise to perturbative corrections at large distances as well as a different running of $g_{\rm NLO}(R)$ with $R$.

% {\bf Where in Sec.~\ref{sec:Results} do we discuss the choice of energy dependence?} Comment has been removed.

For bound states NLO corrections to the LO calculation are computed %perturbatively, 
using standard first-order perturbation theory, i.e., 
\begin{equation}
    \label{eq:nlo_correction}
    B^{(n)}_{\rm NLO} = B^{(n)}_{\rm LO} 
    \left(1 + g_{\rm NLO}(R) \langle\psi^{(n)}_{\rm LO}|\chi|\psi^{(n)}_{\rm LO}\rangle\right)~.
\end{equation}
The value of $g_{\rm NLO}(R)$ is determined here by demanding that the binding energy of the shallowest $S$-wave state, $B^{(6)}_{\rm NLO}$, is fixed to the Temkin-Lamkin result for the pion-hydrogen system, $1.2\times10^{-4}$ a.u.

Phase shifts are similarly computed in first-order perturbation theory according to the distorted-wave Born approximation (DWBA), described in detail for example in Ref. \cite{thompson_nunes_2009}, where the scattering amplitude at on-shell momentum $k=\sqrt{2\mu E}$ is
\begin{equation}
    t^{\rm (NLO)} = t^{\rm (LO)} -\frac{2\mu}{k}\langle \phi^{({\rm LO},-)} | V_{\rm NLO} | \phi^{({\rm LO})}\rangle~,
    \label{tNLO}
\end{equation}
%where the 
with $-$ denoting an outgoing wave. 
%and
%\begin{equation}
%    \label{eq:tLO}
%    t^{({\rm LO})} = V_{\rm LO} + V_{\rm LO} G_0 t^{({\rm LO})}~,
%\end{equation}
%in a short-hand notation where an integral over the momentum
%in the two-body propagator $G_0$ is implicit.
%The phase shifts are obtained from
%\begin{equation}
%    \label{eq:phase_shifts}
%    \delta^{({\rm LO})} = 
%    \tan^{-1}\left(\frac{t^{({\rm LO})}}{1+it^{({\rm LO})}}\right)
%\end{equation}
%at LO. 
% At NLO, we use the same Eq. \eqref{eq:phase_shifts} with 
% $t^{({\rm LO})} \to t^{({\rm NLO})}$.
At NLO, we compute the phase shift perturbatively according to
\begin{equation}
    \label{eq:pert_phase_shift}
    \delta^{({\rm NLO})} = \delta^{({\rm LO})} - \frac{2\mu}{k}\langle \phi^{({\rm LO},-)} | V_{\rm NLO} | \phi^{({\rm LO})}\rangle e^{-2i\delta^{({\rm LO})}}~.
\end{equation}
Alternative but equivalent ways to calculate the NLO scattering amplitude are discussed in App. \ref{app:NLOimplementation}.

%Phase shifts computed at LO and NLO in the following sections, are extracted from the scattering amplitude according to
%\begin{equation}
%    \label{eq:phase_shifts}
%    \delta^{({\rm O})} = \tan^{-1}\left(\frac{t^{({\rm O})}}{1+it^{({\rm O})}}\right)~.
%\end{equation}
% {\bf $t^{(\rm LO)}$ must be defined.} Done.

%\begin{equation}
 %   t \approx t^{(0)} + t^{(1)}~,
%\end{equation}
%and
%\begin{equation}
%    \label{eq:dwba}
%    t^{(1)} = -\frac{2\mu}{k}\langle \phi^{({\rm LO},-)} | V_{\rm NLO} | \phi^{({\rm LO})}\rangle~,
%\end{equation}
%and $|\phi^{({\rm LO})}\rangle$ is the LO scattering state, normalized according to
%\begin{equation}
%    \phi^{({\rm LO})}(r) = e^{i\delta}\sin(kr + \delta)~.
%\end{equation}

\section{Results \label{sec:Results}}

In this section we present numerical results for the scattering of a charged pion on hydrogen, and for the bound states of this system.  As $R$ decreases, $g_{\rm LO}(R)$ becomes very large so it can provide the repulsion necessary to keep $a_0$ and the number of bound states fixed as more of the $1/r^4$ attraction is exposed.
Although working with the fourteen-state branch alleviates the problem, we were unable to find accurate solutions once $\beta_4/R$ became larger than 70.
As we are going to see, this nevertheless is high enough for many conclusions to be drawn about the scope of ID-EFT.

\subsection{Scattering \label{sec:Scattering}}

\begin{figure}
    \centering
    \includegraphics[width=\columnwidth]{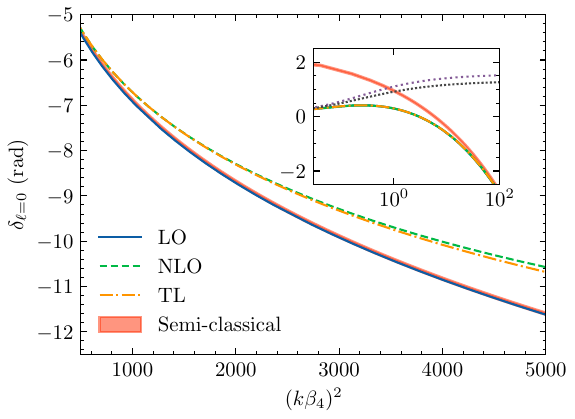}
    \caption{The $S$-wave phase shifts are shown as a function of the dimensionless quantity, $(k\beta_4)^2$.
    The EFT at LO (solid blue line) and NLO (green dashed line) at a cutoff  $R=\beta_4/70$ are compared with the Temkin-Lamkin potential (orange dot-dashed line) \cite{Temkin:1959,Temkin:1961,MM} and a semi-classical approach \cite{Flambaum:1999zza} with $R$ varied between  $\beta_4/45$ and $\beta_4/70$ in the classically forbidden region (red band). 
    ERE results are included in the inset. The purple dotted line represents the $a_0$ term.
    The black, densely dotted line includes the term linear in $k$ as found in Refs.~\cite{OMalley:1961,PhysRev.125.1300}.
    The LO prediction is included in the inset but overlapped by the TL and NLO lines.
    }
    \label{fig:deltas}
\end{figure}

The efficacy of our proposed EFT is first tested in the continuum where we study the LO and NLO $S$-wave phase shifts.
Figure~\ref{fig:deltas} shows the ID-EFT predictions for the phase shifts, alongside the Temkin-Lamkin results, the ERE, and the semi-classical prediction of Ref.~\cite{Flambaum:1999zza}.
We find that the variation of the ID-EFT phase shifts for cutoffs near the breakdown scale is minimal: were we to draw bands of cutoff variation for $\beta_4/R$ above, say, 50, they would be barely visible on the scale of the figure. Therefore, our results are plotted at minimum $R$ (or maximum $\beta_4/R\approx 70$).
We find excellent agreement between ID-EFT and the Temkin-Lamkin phase shifts already at LO over a momentum range that extends well beyond $k \beta_4 \approx 1$, indicating that the breakdown scale of the theory is relatively high.
In other words, the higher momentum range over which this agreement holds suggests that the curvature of the $1/r^4$ potential is a crucial piece of physics in the Temkin-Lamkin phase shifts.
The agreement at lower momenta is due to two factors.
First, the inclusion of the $1/r^4$ interaction allows us to capture physics at the $k\sim 1/\beta_4$ scale.
Second, by fixing $a_0$ at LO, we demand agreement at threshold.
For comparison, the inset of Fig.~\ref{fig:deltas} shows also the first two terms in the ERE from the inverse scattering length $a_0$ and a correction linear in $k\beta_4$~\cite{OMalley:1961,PhysRev.125.1300}. While they approach the TL results for $k\beta_4\ll 1$, ID-EFT captures the sign change of the phase shifts already at LO.
At NLO, as expected, the agreement between ID-EFT and TL improves significantly at larger momenta.

\begin{figure}
    \centering
    \includegraphics[width=\columnwidth]{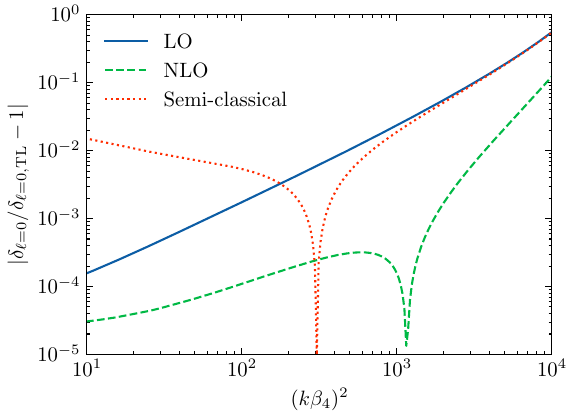}
    \caption{Relative errors with respect to the Temkin-Lamkin $S$-wave phase shifts 
    %\cite{} 
    are shown for the LO (solid, blue line) and NLO (dashed, green line) predictions at maximum $\beta_4/R\approx 70$. The dot-dashed, red line represents the semi-classical prediction \cite{Flambaum:1999zza}. Both scales are logarithmic.
    }
    \label{fig:deltas_relative}
\end{figure}

Figure~\ref{fig:deltas_relative} offers a closer look into the errors in the LO and NLO predictions at minimum $R$.
The most important features 
%of Figure~\ref{fig:deltas_relative} 
are the slopes of the LO and NLO lines --- the rates at which the error grows --- as $k\beta_4$ gets very large.
To interpret this accurately, it is important to keep in mind that the LO and NLO predictions are both dependent on the cutoff, $\Lambda\sim 1/R$, and breakdown scale, $\Lambda_b$.
Where $k$ is greater than the typical momentum scales of the problem, but less than $\Lambda_b$, we expect the $\Lambda$ dependence 
to dominate the errors, as our cutoff is not very large.
This appears in Figure~\ref{fig:deltas_relative} between $(k\beta_4)^2\approx 10^3$ and $10^4$ where the slope of the NLO line is clearly greater than the slope of the LO line.
This is of course by design because at NLO we have suppressed the $\Lambda$ (or $R$) dependence.
Finally, one can expect that these two lines cross above $(k\beta_4)^2\sim10^4$.
This intersection indicates the breakdown scale of ID-EFT in this system,
\begin{equation}
\Lambda_b\sim 100/\beta_4\approx \pi/a_B .
\end{equation}
Thus, the phase-shift results indicate that ID-EFT holds in a region somewhat larger than naively expected.

Bearing this in mind also yields qualitative understanding of other features of Fig.~\ref{fig:deltas_relative}. At low momenta, the NLO prediction is significantly closer to the Temkin-Lamkin result than the LO one is. 
However, the slopes of the LO and NLO deviations on the log-log plot of Fig.~\ref{fig:deltas_relative}
%{fig:lo_nlo_error} 
are similar for $(k \beta_4)^2 \lesssim 10^3$
%$(k \beta_4)^2 < 10^2$, 
even
though NLO is consistently almost an order of magnitude more accurate at these low values of $k$. (Note that 
%even though these values of $k$ are small compared to $1/a_0$, 
$k \beta_4$ is still markedly larger than 1 there, so effective-range theory does not apply.)
The similarity of the slopes 
%of the LO and NLO lines in Fig.~\ref{fig:lo_nlo_error} for $k\beta_4 \gtrsim 10^2$ 
is not coincidental.
Forcing the NLO calculation to reproduce $B_6$ induces an error in the NLO calculation of order $(k \beta_4)^2 2\mu B_{6}/\Lambda_b^2$.
For $k\beta_4 \lesssim 10^3$
%$k\beta_4 \gtrsim 10^3$ 
this effect is larger than the $(k \beta_4)^4$ errors that dominate in the upper end of the EFT's validity range.

This analysis shows that ID-EFT is systematically improvable.
The deviation from the underlying theory---the Temkin-Lamkin potential---is parametrically smaller at NLO than it is at LO.
In contrast with the systematic improvement in ID-EFT, the semi-classical approximation~\cite{Flambaum:1999zza} works as well as LO at high momenta but fails at low momenta.
The semi-classical phase shifts are plotted in Fig.~\ref{fig:deltas} as a band to indicate the variation with respect to $R$ in the range of the classically forbidden region.
It nearly overlaps with the LO curve, but differences are highlighted in the inset of Fig.~\ref{fig:deltas}: its assumptions are clearly not applicable at smaller values of $k$.
Figure~\ref{fig:deltas_relative} reveals that the semi-classical approach describes the Temkin-Lamkin phase shift to better than 2\% once $(k \beta_4)^2 \geq 10$ --- and as long as $k a_B$ remains small. The semi-classical curve crosses the Temkin-Lamkin curve at $(k \beta_4)^2 \approx 3\cdot 10^2$ leading to the dip seen in Fig.~\ref{fig:deltas_relative}, and for $(k \beta_4)^2 \gtrsim 10^3$
%$(k \beta_4)^2 \geq 10^3$ 
it nearly agrees with LO. 

% {\bf which value of $R$ is used for the semiclassical curve in Fig. \ref{fig:deltas_relative}?}

\subsection{Bound States \label{sec:Bound_State_Results}}

Given the success of our description of phase shifts, we now turn to information from the bound-state spectrum, in order to obtain a parallel assessment of the ability of the EFT to capture the energy dependence below threshold for 
%energies 
$|E|\ll \Lambda_b^2/2\mu$.
As $a_0 \approx -2\beta_4$ is negative and not dramatically larger than $\beta_4$, we do not expect that a very shallow bound state 
of size much greater than $\beta_4$ is present in this system, i.e., we anticipate that $B^{(6)}$ is not fine tuned. 

\begin{figure}
    \centering
    \includegraphics[width=\columnwidth]{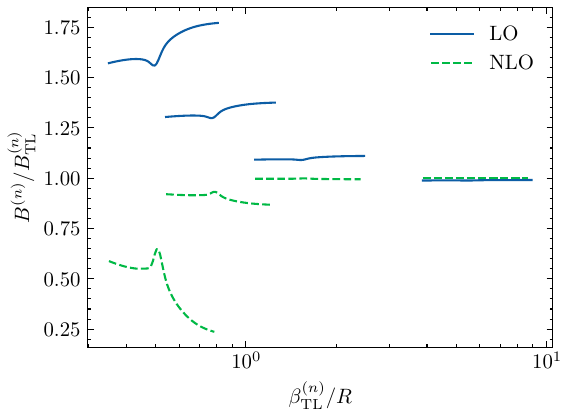}
    \caption{Binding energies from both LO (solid, blue lines) and NLO (dash-dotted, green lines) are shown relative to the Temkin-Lamkin binding energies, $B_{\rm TL}^{(n)}$. The results are plotted against the ratio of length scale associated with each state --- defined in Eq. \eqref{eq:state_length_scale} --- to the short-distance cutoff, $R$. Vertically stacked pairs are shown successively for $n=3$ (leftmost) through to $n=6$ (rightmost). States with $n < 3$ are not shown as it is clear the EFT convergence breaks down for them.}
    \label{fig:lo_nlo_spectrum_R}
\end{figure}

The LO and NLO results for the binding energies $B^{(n)}$ of the four shallowest $S$-wave bound states are shown in Fig.~\ref{fig:lo_nlo_spectrum_R} relative to the Temkin-Lamkin states.
Each state is plotted against  $\beta_{\rm TL}^{(n)}/R$ where
\begin{equation}
\label{eq:state_length_scale}
\beta_{\rm TL}^{(n)} \equiv 1/\sqrt{2\mu B_{\rm TL}^{(n)}}
\end{equation}
is the characteristic size of the $n$th Temkin-Lamkin bound state with binding energy $B_{\rm TL}^{(n)}$.
Because the characteristic size decreases as we go down the spectrum the lines do not cover the same horizontal span even though they are generated with the same $R$ values.
One can see the energies converge as $R$ increases towards $70/\beta_4$, 
but much smaller values of $R$ would be needed to see the deeper states ``flatten out''.

%$B_6$ 
The shallowest, $n=6$, state turns out to be two orders of magnitude deeper than the typical low-energy scale associated with $\beta_4$ in this system, 
\begin{equation}
\epsilon_4 \equiv \frac{1}{ 2\mu\beta_4^2} \approx 1\times10^{-6}~{\rm a.u.}~.
\end{equation}
Nevertheless, we obtain excellent agreement at LO with the Temkin-Lamkin result for this state, which is represented by the rightmost solid, blue line in Fig.~\ref{fig:lo_nlo_spectrum_R}.
This state is still of low energy compared to the energies of states in atomic hydrogen and corresponds to a length scale of approximately 4 a.u.
The fact that this length scale is markedly smaller than $a_0$ demonstrates that our prediction is not a consequence of large-scattering length universality. 
We predict $B^{(6)}$ so well using only $\beta_4$ and $a_0$ as inputs  because we included the attractive $1/r^4$ potential at LO in our EFT and most of the state's wave function extends well outside hydrogen's electron cloud. 
At NLO, we renormalize $g_{\rm NLO}(R)$ to the Temkin-Lamkin binding energy of this state, so agreement  is by construction.

But introducing the  NLO interaction reduces the disagreement between the other binding energies and the Temkin-Lamkin energies. 
The deeper states in Fig.~\ref{fig:lo_nlo_spectrum_R} at smaller 
$\beta_{\rm TL}^{(n)}/R$  values are an excellent visualization of how the theory scales with energy.
$B^{(5)}$, the fifth excited state, second pair of lines from the right, is captured to within $\approx$10\% at LO and $\approx$1\% at NLO.
Moving downward in the spectrum, where the pion's wave function has more overlap with the hydrogen atom, the NLO error for $B^{(4)}$ and $B^{(3)}$ grows systematically larger.
This is a natural outcome in ID-EFT. 
We are fixing the scattering amplitude at threshold and continuing it to negative energies to find poles, including the analyticity properties implied by the $1/r^4$ potential in that continuation.
In fact, by the time we reach $B^{(3)}$, NLO is not an improvement --- both LO and NLO are off by 50\%.
This indicates that the series does not converge, which is why the ground and first two excited states are not shown in Fig.~\ref{fig:lo_nlo_spectrum_R}.
As the continuation is made over a bigger energy range our prediction becomes less accurate.

In order to extrapolate our results to the $R \rightarrow \infty$ limit we assume that the effects associated with finite $R$ can be accounted for via an expansion in $R/\beta_4$. 
Therefore, we rely on the assumed convergence behavior of observables close to the renormalization point to extract asymptotic estimates. We expand
\begin{equation}
    \label{eq:convergence}
    \mathcal{O}(R) = \mathcal{O}_\infty\left[ 1 + \sum_{m=1}^\infty c_m\left(\frac{R}{\beta_4}\right)^m\right]~,
\end{equation}
with $\mathcal{O}_\infty$ being the asymptotic result for the LO or NLO energy of these bound states in ID-EFT and $c_m$ the coefficients of the expansion.
In the leading-order case the  ID-EFT results for all seven binding energies are found by fitting the first coefficient of the expansion \eqref{eq:convergence} and then reporting only ${\mathcal O}_\infty$ in Table~\ref{tab:binding_energies}. At NLO the situation is more complicated: both the $m=1$ and $m=4$ term are needed to accurately fit the data. The rationale for this fit function is explained in Appendix~\ref{app:nlo_fit}.

The NLO results show convergence to the TL results for the shallowest three states, but for the lowest three states NLO repulsion is so strong that they are no longer bound.
(This further supports the use of the 14-state branch in our local regulator scheme.)
This reordering of the states at NLO occurs already for the $n=3$ state: its repulsive NLO correction is so large that it renders $B_{\rm NLO}^{(3)}$  smaller than $B_{\rm NLO}^{(4)}$.
The matrix element in Eq.~\eqref{eq:nlo_correction} is clearly no longer a perturbation for $n\le 3$.

\begin{table}[tb]
    \centering
    \begin{tabular}{c|c|c|c}
        $n$ & $B_{\rm LO}$ (a.u.) & $B_{\rm NLO}$ (a.u.)  & $B_{\rm TL}$ (a.u.) \\
        \hline
        \hline
        6 & 1.19e-04 & 1.20e-04 & 1.20e-04 \\
        \hline
        5 & 1.75e-03 & 1.56e-03 & 1.56e-03 \\
        \hline
        4 & 8.59e-03 & 5.39e-03 & 6.12e-03 \\
        \hline
        3 & 2.72e-02 & 4.91e-03 & 1.47e-02 \\
        \hline
        2 & 6.84e-02 & -3.02e-02 & 2.74e-02 \\
        \hline
        1 & 1.49e-01 & -1.84e-01 & 4.42e-02 \\
        \hline
        0 & 2.98e-01 & -6.95e-01 & 6.48e-02 \\
        \hline
    \end{tabular}
    \caption{Asymptotic ($R\rightarrow\infty$) results for LO and NLO binding energies compared to the Temkin-Lamkin spectrum. LO results are obtained from a fit to Eq. \eqref{eq:convergence}. Details about the NLO fit are given in Appendix~\ref{app:nlo_fit}. Negative values indicate that the state is not bound.}
    \label{tab:binding_energies}
\end{table}
% percent errors update 2023/07/07

\begin{figure}
    \centering
    \includegraphics[width=\columnwidth]{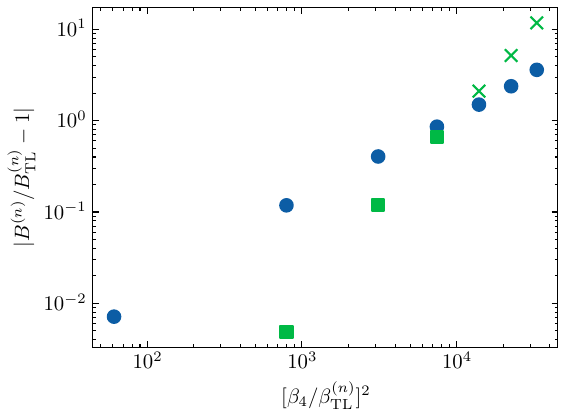}
    \caption{The relative error between the Temkin-Lamkin binding energies, $B_{\rm TL}$, and those obtained with ID-EFT. LO results are shown as blue circles. NLO results are shown in green --- squares indicate bound states while x's indicate that the NLO correction pushes the state above the continuum.}
    \label{fig:lo_nlo_error}
\end{figure}

The error of these binding energies relative to the Temkin-Lamkin result is shown in Fig.~\ref{fig:lo_nlo_error}.
It ranges from 1\% for the sixth excited state, to 12\% for the fifth excited state, to a factor of 4 for the ground state.
It is notable that the LO error grows linearly with the energy of the bound state: the slope is $\approx 1$. The NLO interaction removes this error but leaves errors quadratic in energy, and indeed at NLO the slope is $\approx 2$.

Thus, we find evidence that the three shallowest states are within the regime of validity of ID-EFT, even though their binding energies vary by a factor of $\approx 100$. The breakdown binding energy $\sim 10^{-2}$ a.u. inferred from where the LO and NLO trends intersect in Fig.~\ref{fig:lo_nlo_error} implies $\Lambda_b\sim 100/\beta_4$. This value is in good agreement with the determination from scattering.

\section{Conclusions \label{sec:Conclusions}}

We presented an effective field theory, ID-EFT, to describe the low-energy scattering of a heavy charged particle on a neutral atom, and the associated shallowest bound states. This EFT captures at leading order the physics of the long-range, but singular attractive, potential created by the atom's polarization. Renormalization requires at LO also a short-range interaction that is fixed by one datum. At next-to-leading order a second short-range interaction, determined by a second datum, systematically improves results for observables till momenta reach the breakdown scale.

We illustrated the workings of the EFT above and below threshold when the charged particle is a pion and the atom is hydrogen. We took as data results of the Temkin-Lamkin potential --- the scattering length at LO and the shallowest binding energy at NLO --- and compared the EFT outcomes with the exact results for other predictions using the same potential. Because the scattering length is not particularly large nor the shallowest state particularly shallow in the scale set by the long-range potential, there is no fine tuning in this system and the EFT goes well beyond the effective range expansion. 

We found a momentum breakdown scale somewhat larger (by a factor $\approx \pi$) than the inverse of the Bohr radius, $1/a_B$. For smaller momenta, phase shifts are well described at LO and the description improves systematically at NLO. The three shallowest bound states are also better reproduced at NLO than at LO. For larger momenta, the pion probes the inside of the atom; the atom can no longer be treated as a single unit. The four lowest-lying states of the pion-hydrogen ion have sizes somewhat smaller than $a_B$ and are outside the regime of validity of the EFT. 

Although here we used the Temkin-Lamkin potential as an example, ID-EFT offers a simple way to account for the long-range properties of this type of system without requiring detailed knowledge of the dynamics inside the atom. ID-EFT can be applied to other heavy charged particles and/or atoms and compared to data and/or other calculations where atomic structure is taken into account. We expect to find a similar convergence pattern, although details will depend on the values of leading-order parameters and the breakdown scale.

\section*{Acknowledgements}

We thank the Kavli Institute for Theoretical Physics for accommodating us and facilitating the initial steps of this work during the program ``Living Near Unitarity".
We appreciate the useful discussions we had with Jaume Carbonell and Rimantas Lazauskas during and since that program. 
This research was supported in part by the National Science Foundation under Grant Nos. NSF PHY-1748958 and OAC-2004601 (CSSI program, BAND
collaboration), and by the US Department of Energy under 
%contact 
award numbers no. DE-FG02-93ER40756 and DE-FG02-04ER41338.
%(UvK). 

\appendix

\section{Accurate Calculations of the Scattering Length in the Presence of a $1/r^4$ Potential}
\label{app:scattering_length}

The zero-energy solution to the reduced radial Schr\"odinger equation for a finite-range potential goes asymptotically like
\begin{equation}
    \label{eq:zero_energy_solution}
    u(r) \propto 1 - r/a_0~.
\end{equation}
In general, the solution can be calculated numerically and the long-range portion of the wave function can be fit to a straight line such that the slope and intercept give an accurate and stable estimate of the scattering length.
However, once a ``long-range'' potential is introduced, the tail of the interaction can make this extraction slow to converge.

In order to overcome this challenge, we derive the so-called ``infrared corrections'' perturbatively assuming that the $1/r^4$ potential is weak at large distances.
The exact solution becomes a sum
\begin{equation}
    u(r) = \sum_{i=0}^\infty u^{(i)}(r)~.
\end{equation}
The first-order correction at zero energy is then
\begin{equation}
    \label{eq:ir_correction}
    -\frac{d^2}{dr^2}u^{(1)}(r) = \frac{\beta_4^2}{r^4} u^{(0)}(r)~,
\end{equation}
where $u^{(0)}(r)$ is taken to be Eq.~\eqref{eq:zero_energy_solution} up to an overall factor.
After integration, we obtain
\begin{equation}
    \label{eq:u1_correction}
    u^{(1)}(r) = \frac{\beta_4^2}{2a_0}\left( \frac{1}{r} - \frac{a_0}{3r^2}\right)~,
\end{equation}
and a better approximation for $u(r)$ (again, up to an overall factor),
\begin{equation}
    \label{eq:inverse_fourth_0_energy_wf}
    u(r) \approx 1 - \frac{r}{a_0} + \frac{\beta_4^2}{2a_0}\left( \frac{1}{r} - \frac{a_0}{3r^2}\right)~.
\end{equation}
With this corrected form of the zero-energy solution, we are able to fit the coefficients of
\begin{equation}
    \label{eq:inverse_fourth_0_energy_form}
    u(r) = b_0 + b_1r + b_{-1}/r + b_{-2}/r^2~,
\end{equation}
and reliably extract $a_0=-b_0/b_1$ at much lower $r$.
Additionally, we are able to compare the fit to the predicted coefficients of the $1/r$ and $1/r^2$ terms where
\begin{eqnarray}
    \frac{\beta_4^2}{2a_0} & = \frac{b_{-1}}{b_0}~, \\
    \frac{\beta_4^2}{6} & = \frac{b_{-2}}{b_0}~.
\end{eqnarray}
We find excellent agreement between the fit results and the analytical predictions.

\section{Local Branches}
\label{app:branches}

Several aspects of renormalization studies depend strongly on the ability to numerically approximate the limit $\beta_4/R\rightarrow\infty$.
In local systems characterized by singular interactions, as $R$ decreases and more of the singular well is exposed, the strength of the repulsive counterterm grows quickly.
For these same local systems, there are an infinite number of ``branches'', each corresponding to a unique number of bound states---as noted in Ref.~\cite{Beane:2001}.
In order to achieve the practical $\beta_4/R\rightarrow\infty$ limit for the seven states mimicking the Temkin-Lamkin states, we chose to leverage this option and work on the $n=14$ branch.

\begin{figure}[t]
    \centering
    \includegraphics[width=\columnwidth]{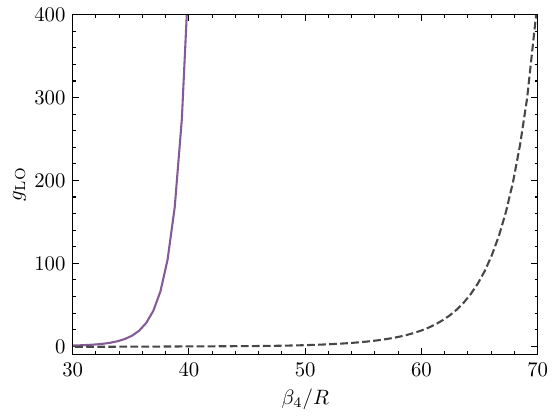}
    \caption{The running of the LO coupling for the $n=7$ (solid, purple) and $n=14$ branches (dashed, black).}
    \label{fig:rgflows}
\end{figure}

The advantages of this choice are highlighted in Fig.~\ref{fig:rgflows}, where $g_{\rm LO}^{(n=7)}$ begins to increase rapidly at $R\approx\beta_4/40$, but $g_{\rm LO}^{(n=14)}$ does not reach the same value until $R\approx\beta_4/70$. 
Using this deeper branch, 
shallower states reach the asymptotic regime --- where Eq.~\eqref{eq:convergence} is valid --- faster.
On the $n=14$ branch, the seven states of interest are now the shallowest states, and we are able to capture far more of the asymptotic behavior.

\section{NLO Implementation Comparison}
\label{app:NLOimplementation}

The NLO amplitude can be written in different forms, some of which we briefly compare here. In all cases, we define 
\begin{equation}
t^{(\rm NLO)} \equiv t^{(\rm LO)} + \delta t~,
\label{tNLOprime}
\end{equation}
and the LO $t$ matrix is given by Eq.~\eqref{eq:tLO}.

In the first method, presented in Sec. \ref{sec:Theory_NLO}, 
\begin{equation}
    \delta t = -\frac{2\mu}{k}\langle \phi^{({\rm LO},-)} | V_{\rm NLO} | \phi^{({\rm LO})}\rangle~.
    \label{deltat1}
\end{equation}
From the relation between scattering wave function and the scattering amplitude, we can write instead
\begin{equation}
\delta t = (1 + t^{({\rm LO})}G_0)V_{\rm NLO}(G_0t^{({\rm LO})} + 1)~.
\label{deltat2}
\end{equation}
This form is used, for example, in Ref.~\cite{LongYangSinglet}. 
By substituting Eq.~\eqref{eq:tLO} in Eq.~\eqref{deltat2},
\begin{eqnarray}
\delta t 
%&=& (1 + V_{\rm LO}G_0 + V_{\rm LO} G_0 t^{({\rm LO})}G_0)
%V_{\rm NLO}(G_0t^{({\rm LO})} + 1)
%\nonumber\\
&=& V_{\rm NLO} + V_{\rm NLO} G_0 t^{({\rm LO})} 
\nonumber\\
&& + V_{\rm LO} G_0 (1+  t^{({\rm LO})}G_0)
V_{\rm NLO}(G_0t^{({\rm LO})} + 1)
\nonumber\\
&=& V_{\rm NLO} + V_{\rm NLO} G_0 t^{({\rm LO})} 
+ V_{\rm LO} G_0 \; \delta t~,
\label{deltat3}
\end{eqnarray}
which is the form used in Refs.~\cite{Vanasse2013,Koenig_2017}.
We have checked numerically that these three equivalent forms for $\delta t$ indeed give the same result.

Since the calculation of $\delta t$ respects unitarity only perturbatively there are then  different ways to express the relation between $t^{({\rm NLO})}$ and $\delta^{({\rm NLO})}$. Starting with the standard (unitary) relationship between the scattering amplitude and the phase shift,
\begin{equation}
    \label{eq:t_nlo}
    t^{({\rm NLO})} = \frac{1}{2i}\left( e^{2i\delta^{({\rm NLO})}} - 1 \right)~,
\end{equation}
we write $\delta^{({\rm NLO})} = \delta^{({\rm LO})} + \epsilon$ and expand in small $\epsilon$, ignoring $\mathcal{O}(\epsilon^2)$ terms and higher. This yields
\begin{equation}
    \label{eq:t_nlo_approx}
    t^{({\rm NLO})} \approx \frac{1}{2i} \left[ e^{2i\delta^{({\rm LO})}}(1+2i\epsilon) - 1 \right]~.
\end{equation}
Simplifying, we get
\begin{equation}
    t^{({\rm NLO})} = t^{({\rm LO})} + \epsilon ~e^{2i\delta^{({\rm LO})}}~,
\end{equation}
where
\begin{equation}
    t^{({\rm LO})} = \frac{1}{2i}\left( e^{2i\delta^{({\rm LO})}} - 1 \right)~
\end{equation}
and the NLO correction to the scattering amplitude is related to the NLO piece of the phase shift by
\begin{equation}
    \label{eq:delta_nlo_pert}
    \epsilon = 
    e^{-2i\delta^{({\rm LO})}} \delta t~.
\end{equation}
This relation is strictly perturbative, in the sense that all quantities are computed to NLO accuracy and not further.

\begin{figure}[t]
    \centering
    \includegraphics[width=\columnwidth]{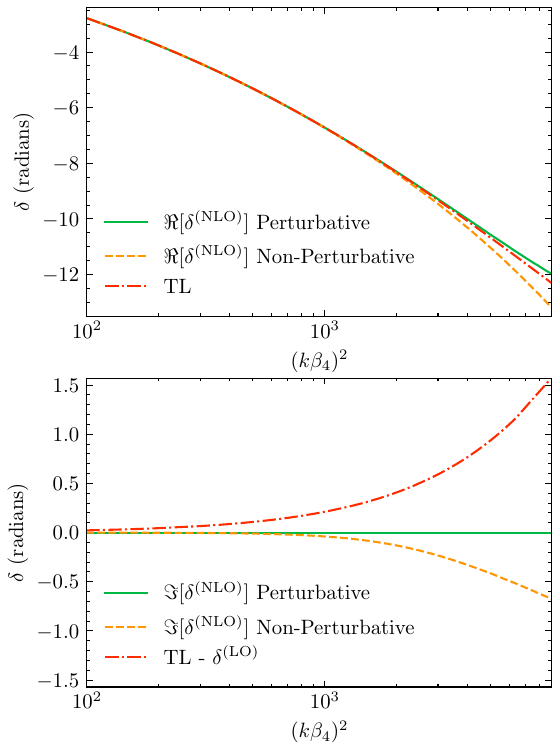}
    \caption{(Upper panel) The real parts of the NLO phase shifts computed using Eq.~\eqref{eq:delta_nlo_pert} (solid, green line) and Eq.~\eqref{eq:t_nlo} (dashed, orange line). The TL results are shown as a dash-dotted, red line. (Lower panel) Imaginary components of the NLO phase shifts (colors and styles are the same as in the upper panel). The (strictly real) difference between the TL results and the LO phase shifts is indicated with a dash-dotted, red line.}
    \label{fig:nlo_phase_shift_comparison}
\end{figure}

If instead one computes $\delta^{({\rm NLO})}$ from $\delta t$ ``non-perturbatively'',  via Eq.~\eqref{eq:t_nlo}, thereby assuming that unitarity remains strictly valid at NLO, two key differences emerge.
First, the real part of the phase shift is less accurate, as seen in the upper panel of Fig.~\ref{fig:nlo_phase_shift_comparison}.
Second, an imaginary component accumulates at higher momenta --- shown explicitly in the lower panel Fig.~\ref{fig:nlo_phase_shift_comparison}.
Consistently computing the scattering amplitude and phase shifts perturbatively produces not only formally correct results but more accurate predictions.

Note that the imaginary component of the phase shift accumulated in such a calculation, although unphysical, does provide insight into the expected size of the NNLO correction.
The first term omitted in the expansion of $\exp(2 i \epsilon)$ in Eq.~\eqref{eq:t_nlo_approx} is real and $\mathcal{O}(\epsilon^2)$.
If we consider also the prefactors in Eq.~\eqref{eq:t_nlo_approx}, we conclude that this is the piece of Eq.~\eqref{eq:t_nlo} that generates the leading piece of the imaginary part of the non-perturbative phase shift, and hence $\Im(\delta^{\rm (NLO)})$ should also be $\mathcal{O}(\epsilon^2)$.
In the lower panel of Fig.~\ref{fig:nlo_phase_shift_comparison}, we see in the context of the $\mathcal{O}(\epsilon)$ difference between the TL phase shifts and our LO phase shifts that this conclusion is well-supported.

Retaining $\Re (\delta^{({\rm NLO})})$ from Eq. \eqref{eq:t_nlo}, and simply ignoring $\Im (\delta^{({\rm NLO})})$, accounts for some effects beyond NLO. But this is an incomplete NNLO calculation. The line this produces in Fig. \ref{fig:deltas_relative} has  approximately the same slope as the NLO line, which is based on Eq.~\eqref{eq:delta_nlo_pert}. However, as $\Re(\delta^{({\rm NLO})})$ calculated ``non-perturbatively'' gives worse results (compared to the TL phase shifts) than the consistent perturbative calculation, the line this produces in Fig. \ref{fig:deltas_relative} is displaced up, towards the LO curve. It then intersects the LO curve at a smaller momentum, leading to an underestimate of the breakdown scale. This is just a new example of a well-known phenomenon: including a partial subset of small corrections does not necessarily improve the result. For another example of the same type, see Fig. 5 of Ref. \cite{Stetcu:2010xq}, and for a more dramatic example, with far-reaching consequences, Ref. \cite{Phillips:1996ae}.

\section{$R$ Dependence of the NLO Binding Energies}
\label{app:nlo_fit}

%\begin{itemize}
%    \item Eq.~\eqref{eq:convergence} works for LO but not NLO
%    \item Matrix elements in Eq.~\eqref{eq:nlo_correction} can be expanded in $R^2$ (parity)
%        \subitem Maybe $R^4$ because of the form of the short-distance, local regulator
%    \item NLO binding energies inherit $R$ term from $B_{\rm LO}(R)$
%    \item So we anticipate $R + R^2$ or $R + R^4$ dependence.
%    \item Dominant $R$ dependence in region where we have fit data is $R^4$
%        \subitem $R^2$ fit clearly does not work
%        \subitem Why $R^4$ and not $R^2$ is deferred to future work.
%    \item After that $R^4$ dependence is subtracted remaining dependence looks very linear in region where we have data; and linear term has opposite sign to $R^4$ term, and same sign as at LO for shallow states
%        \subitem Residuals of $R + R^4$ fit exhibit no systematic (power-law dependence)
%        \subitem Remaining uncertainty is within tolerance of reported values
%    \item Coefficient of $R^4$ term is kind-of natural. Still some systematic dependence we don't understand.
%    \item Coefficient of $R$ term is smaller than LO coefficient of $R$ term for states with binding momentum similar to the momentum of the renormalization point.
%    \item Coefficient of $R$ term grows as binding momentum grows, and grows faster with binding momentum than LO term does. Still some systematic dependence we don't understand.
%\end{itemize}

Extracting asymptotic binding energies at LO is relatively straightforward, but it turns out to be markedly more complicated at NLO. 
First we observe that the NLO binding energy formula, Eq.~\eqref{eq:nlo_correction}, contains two different $R$-dependent terms.
The first is $B_{\rm LO}(R)$.
The second includes a matrix-element ratio when the NLO renormalization condition
\begin{equation}
    \label{eq:g_nlo}
    g_{\rm NLO}(R) = \left(\frac{B^{(6)}_{\rm TL}}{B^{(6)}_{\rm LO}} - 1 \right) \frac{1}{\langle\psi^{(6)}_{\rm LO}|\chi|\psi^{(6)}_{\rm LO}\rangle}
\end{equation}
is inserted.
$B_{\rm LO}^{(n)}(R)$ is well-described with a straight line, i.e. without including any of the $m>1$ terms in Eq.~\eqref{eq:convergence}.
The matrix-element ratio is more complicated: we expect it to be even in $R$ because of analyticity in the regulator parameter. But whether the non-linear piece of the NLO binding energy is proportional to $R^2$ (simplest dependence) or $R^4$ (which is the leading $R$ dependence of the regulator) is not immediately apparent. Consequently, 
we have tried several approaches to fit the NLO binding energies and record them here for posterity.

We require two features of our NLO binding energy analysis.
First, we expect that
the NLO renormalization procedure ought to reduce the linear $R$ dependence for states where the binding momentum is not far from the additional fixed point, $\gamma^{(6)}$: for these states the coefficient of the linear-in-$R$ term will be smaller than those found when fitting the LO binding energies.
Second, we expect that the EFT fails {\textit systematically}.
The NLO corrections ought to increase with a positive power of the binding momentum.

\begin{figure}[t]
    \centering
    \includegraphics[width=\columnwidth]{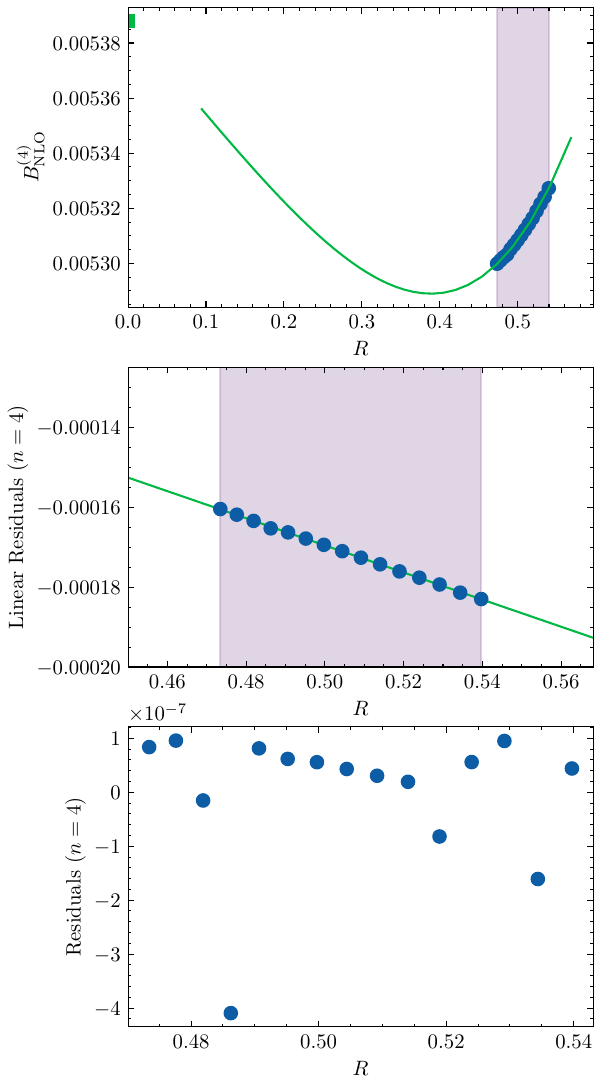}
    \caption{Results of a fit to Eq.~\eqref{eq:convergence} where only $m=1$ and $m=4$ terms are included. (Top panel) Binding energy as a function of the cutoff distance, $R$. Both quantities are in atomic units. Data are indicated as blue circles. The fit is represented with a solid, green line. The asymptotic value of the binding energy is indicated on the y axis as a green square. Purple regions highlight the range of $R$ over which the fit was conducted. (Middle panel) Data and fit results for the linear term of our fit are shown with the asymptotic binding energy and the best-fit quartic $R$ dependence subtracted. (Bottom panel) The residuals of our $R + R^4$ fit. Note that the $y$-scale in the bottom panel is expanded roughly 100 times compared to the top panel. Note also that the difference $x$-axis range gets progressively smaller as one moves from top to middle to bottom panel.}
    \label{fig:nlo_binding_energies}
\end{figure}

Figure~\ref{fig:nlo_binding_energies} displays the steps we took to understand the $R$ dependence of the binding energy of the fourth excited state in our NLO calculation. 
The upper panel of Fig.~\ref{fig:nlo_binding_energies} shows the $R$ dependence of this quantity is not exclusively linear.
A successful fit to Eq.~\eqref{eq:convergence} is shown together with the data.
While the ability of the fit to describe the data is clear (and highlighted in the bottom panel), it seems odd at first glance that data that is decreasing as $R$ decreases leads to an asymptotic value that is larger than any value in the data set. The minimum, and subsequent change in derivative, of $B_{\rm NLO}^{(4)}$ with respect to $R$, is due to the linear term in the fit assuming a dominant role, a role that it does not have for the $R$ values at which the fit is performed.

The middle panel of Fig.~\ref{fig:nlo_binding_energies} shows that we can confidently extract the linear coefficient in the region where we have data on the bound state energy. In this panel we have subtracted the asymptotic value and quartic term from the data, leaving only the linear dependence. A linear fit is clearly an excellent description of the residuals: the linear dependence that dominates at small $R$ is indeed present in the data even where the quartic term is more significant.

The bottom panel of Fig.~\ref{fig:nlo_binding_energies} then shows the residuals once the linear-in-$R$ term in our fit is subtracted. The residuals show no systematic trend with $R$ and are of order $10^{-7}$ in atomic units. Any remaining $R$ dependence will therefore not affect the extrapolation at the level of accuracy we are quoting in this paper.

To be thorough, we also fitted out data using $R+R^2$ and $R+R^2+R^4$ forms.
While these approaches did align with the data in the region $0.467 < R < 0.54$, they also produced coefficients of the linear-in-$R$ term that were larger than the LO value of the same coefficient for states with $n\ge 4$, i.e., states where including NLO corrections and renormalizing to $B_6$ should have decreased the linear dependence on $R$. These fits also yielded
asymptotic values of the binding energies that had no systematic trend with $n$, and showed clear signs of over-fitting. We defer the question of why there is no $R^2$ term present in the function $B_{\rm NLO}(R)$ to future work, only commenting here that we do not believe this behavior will prevail for all regulators.

\bibliography{refs}% Produces the bibliography via BibTeX.

\end{document}